\documentclass[pra,twocolumn,showpacs,superscriptaddress,byrevtex]{revtex4}

\usepackage{graphicx}
\usepackage{dcolumn}
\usepackage{amsmath}
\usepackage{epsfig}
\usepackage{bm}
\usepackage{amssymb}

\newcommand{\ket}[1]{\ensuremath{|\,{#1}\,\rangle}}
\newcommand{\bra}[1]{\ensuremath{\langle\,{#1}\,|}}
\newcommand{\braket}[2]{\ensuremath{\langle\,{#1}\,|\,{#2}\,\rangle}}

\newcommand{\lsub}[1]{\ensuremath{_{_{\!\scriptstyle #1}}}}
\newcommand{\ce}[1]{\ensuremath{\mathcal{#1}}}

\newcommand{\st}{\scriptstyle}

\newcommand{\itg}[1]{\ensuremath{\int\!\!d{#1}\!\!}}
\newcommand{\itgf}[1]{\ensuremath{\int\!\!d{#1}\,}}

\newcommand{\sinc}{\ensuremath{\mbox{\hspace{1.3pt}sinc}\,}}

\begin{document}


\title{Propagation of spatially entangled qudits through free space}

\date{\today}

\author{G. Lima}
\author{Leonardo Neves}
\author{Ivan F. Santos}
\affiliation{Departamento de F\'{\i}sica, Universidade Federal de
Minas Gerais, Caixa Postal 702, Belo~Horizonte,~MG 30123-970,
Brazil.}

\author{J. G. Aguirre G\'omez}
\affiliation{Center for Quantum Optics and Quantum Information,
Departamento de Fisica, Universidad de Concepci\'on,  Casilla
160-C, Concepci\'on, Chile.}

\author{C. Saavedra}
\affiliation{Center for Quantum Optics and Quantum Information,
Departamento de Fisica, Universidad de Concepci\'on,  Casilla 160-C,
Concepci\'on, Chile.}

\author{S. P\'adua}
\email{spadua@fisica.ufmg.br} \affiliation{Departamento de
F\'{\i}sica, Universidade Federal de Minas Gerais, Caixa Postal
702, Belo~Horizonte,~MG 30123-970, Brazil.}



\begin{abstract}

We show the propagation of entangled states of high-dimensional
quantum systems. The qudits states were generated using the
transverse correlation of the twin photons produced by spontaneous
parametric down-conversion. Their free-space distribution was
performed at the laboratory scale and the propagated states
maintained a high-fidelity with their original form. The use of
entangled qudits allow an increase in the quantity of information
that can be transmitted and may also guarantee more privacy for
communicating parties. Therefore, studies about propagating
entangled states of qudits are important for the effort of
building quantum communication networks.

\end{abstract}

\maketitle


\section{Introduction}

Most of the applications in quantum communication, like
teleportation \cite{Bennett} and quantum cryptography
\cite{Ekert}, rely on the assumption that the communicating
parties are capable to transmit entangled particles between
themselves. Because of the practical potential that an
implementation of these applications over distant locations could
have, the propagation of entangled states of qubits have been
theme of recent studies. The first remarkable work used
optical-fibers links to send energy-time entangled qubits for
receivers separated by more than $10 \, Km$ \cite{Gisin1}. A test
of Bell inequality \cite{Bell} showed that the two-photon state
was still an entangled state and it was the first evidence that
quantum correlations could be maintained over significant
distances. Another interesting work was about the free-space
distribution of polarization entangled qubits through the
atmosphere \cite{Zeilinger1}. As it was emphasized in that paper,
``... one of the benefits of a free-space distribution of quantum
entanglement is the possibility of bridging large distances by
additional use of space infrastructure...". Observers were
separated by $600 \, m$ and the quality of the entanglement of the
propagated state had been guaranteed by a violation of Bell's
inequality by more than four standard deviations.

Even though promising new experiments have had success propagating
entangled qubits over farther distances \cite{Gisin2,Zeilinger2},
it has been demonstrated theoretically by E. Waks \emph{et al.}
\cite{Waks}, that due to channel losses and dark counts, the
communication length cannot surpass the order of $100 \, km$ while
using entangled photons and joint measurements. For this reason we
believe that the control of the technique to create and transmit
entangled photons lying in a D-dimensional ($D \geq 3$) Hilbert
space, will be a crucial step in the near future. They allow an
increase in the quantity of information that can be transmitted
per pair shared and will then require less effort of quantum
repeaters when transmitting information in a global scale. Another
advantage is that the use of entangled qudits may increase the
security of the entanglement based quantum cryptography protocols
against certain types of eavesdropping attacks \cite{Durt}.

In this work, we report the experimental free-space propagation of
two entangled 4-dimensional qudits or ququarts. Following the
studies developed at references \cite{Leonardo,Glima,Boyd}, the
ququarts entangled state was generated by using the transverse
spatial correlation of the photon pairs (\emph{biphotons})
produced by spontaneous parametric down-conversion (SPDC) and two
four-slits sets, where they were transmitted through. The
propagation was performed at laboratory scale and the propagated
state observed had a high-fidelity with its original form. The
presence of interference when the two photons are detected in
coincidence is used as an experimental measurement for showing
that the state of the propagated ququarts is entangled and an
evidence of the good quality of the entanglement is discussed.


\section{Theory}

In Ref. \cite{Leonardo}, it was showed that the state of the
biphotons when they are transmitted through generic apertures is

\begin{equation}     \label{psislits}
\ket{\Psi}=\itg{q_{1}}\itg{q_{2}} \,\,\,
  \ce{F}(q_{1},q_{2})\ket{1q_{1}}\ket{1q_{2}},
\end{equation} with the biphoton amplitude given by

\begin{eqnarray}   \label{F}
\ce{F}(q_{1},q_{2}) & \propto \!\!\!\! & \itg{x_{1}}\itg{x_{2}}
\,\,\, e^{i\frac{k}{8z_{A}}(x_{2}-x_{1})^{2}}
e^{-i(q_{1}x_{1} + q_{2}x_{2})} \nonumber \\[2mm]
&  &  \times A_{1}(x_{1}) A_{2}(x_{2})
W\!\bm{(}{\st\frac{1}{2}}(x_{1}+x_{2});z_{A}\bm{)},
\end{eqnarray} where $q_{j}$ and $x_{j}$ are the wave vector and position
transverse components, respectively, of the down-converted photons
in modes $j=1,2$. $A_{j}(x_{j})$ is the transmission function of
the aperture in mode $j$ and $W(\xi ;z_{A})$ is the pump beam
transverse field profile at the plane of the apertures
($z=z_{A}$).

We consider the apertures where the twin photons are sent through
as two identical four-slits. The separation between two
consecutive slits is $d$ and $a$ is the slits half width. If the
pump beam transverse profile, $W(\xi ;z_{A})$, is non-vanishing
only at a small region of space, it can be experimentally
demonstrated that Eq.~(\ref{psislits}) becomes \cite{Glima}

\begin{equation}        \label{qudits}
\ket{\Psi} = \frac{1}{2} \sum_{l=-\frac{3}{2}}^{\frac{3}{2}}
          e^{ik\frac{d^{2}l^{2}}{2z_{A}}} \;
          \ket{l}\lsub{1} \otimes \ket{-l}\lsub{2},
\end{equation}
where

\begin{equation}      \label{base}
\ket{l}\lsub{j} \equiv \sqrt{\frac{a}{\pi}}
                \itgf{q_{j}} e^{-iq_{j}ld}\sinc(q_{j}a)\ket{1q_{j}}.
\end{equation}
The $\{\,\ket{l}\lsub{j}\}$ states represent the photon in mode
$j$ transmitted by the slit $l$ and satisfy
$\lsub{j}\braket{l}{l'}\lsub{j}=\delta_{ll'}$. The two-photon
state in Eq.~(\ref{qudits}) is a maximally entangled state of two
ququarts, where we can see that, when the photon in mode 1 is
transmitted by the slit $l$ the photon in mode 2 will pass through
the symmetrically opposite slit $-l$.

\begin{figure}[tbh]
\begin{center}
\includegraphics[width=0.30\textwidth]{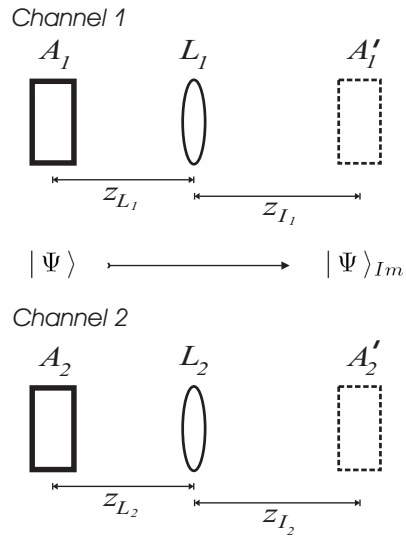}
\end{center}
\caption{\label{fig:channels} Channels for the free-space
propagation. $A_1$ and $A_2$ are generic apertures. $L_1$ and
$L_2$ are convergent lenses with focal lengthes $f_1$ and $f_2$,
respectively. $A'_1$ and $A'_2$ are the images of the apertures.}
\end{figure}

Now we want to show that the state of Eq.~(\ref{qudits}) can be
propagated at the free-space. The biphotons propagation will be
through two independent channels which have distinctly lenses with
focal length $f_1$ and $f_2$ (Fig.~(\ref{fig:channels})). Theses
lenses are placed at a distance $z_{L_i}$ from their respective
apertures. We calculated the two-photon state transmitted by
generic apertures and propagated through these channels to the
planes of the image formation ($z_{I_i}$).

For simplicity, the conditions used for image formation are
$z_{I_i} - z_{L_i} = z_{L_i} - z_{A} = 2 f_i$. To obtain the image
state, a general form for it must be assumed

\begin{equation}     \label{psiimage}
\ket{\Psi}_{Im}=\itg{q_{1}}\itg{q_{2}} \,\,\,
  \ce{I}(q_{1},q_{2})\ket{1q_{1}}\ket{1q_{2}}.
\end{equation}

Calculating the amplitude of the coincidence detection
\cite{Mandel} of the biphotons at the planes of the image
formation using two different methods, we could establish the form
of $\ce{I}(q_{1},q_{2})$. The first amplitude's calculus was done
considering the state of Eq.~(\ref{psislits}) and the
electric-field operators describing the evolution of the photons
through their channels. The second method used the state of
Eq.~(\ref{psiimage}) and the expression for the electric-field
operator at the point of image formation. Matching their results
we obtained

\begin{eqnarray}     \label{psiimage2}
\ce{I}(q_{1},q_{2})&\propto & \itg{q'_{1}}\itg{q'_{2}}\,\,\,
  \ce{F}(q'_{1},q'_{2}) \nonumber \\
  & &\times e^{-if_{1}(q_{1}+q'_{1})^{2}/{2k}} e^{-if_{2}(q_{2}+q'_{2})^{2}/{2k}}.
\end{eqnarray}

When the apertures from which the twin photons were transmitted by
are two identical four-slits, described above, the function
$\ce{F}(q_{1},q_{2})$ will be given by \cite{Leonardo}

\begin{equation}
\ce{F}(q_{1},q_{2}) \propto \sum_{l=-\frac{3}{2}}^{\frac{3}{2}}
e^{ik\frac{d^{2}l^{2}}{2z_{A}}} e^{-iq_{1}ld}\sinc (q_{1}a)
e^{iq_{2}ld}\sinc (q_{2}a). \label{eq:fend}
\end{equation}

Thus, inserting Eq.~(\ref{eq:fend}) into Eqs.~(\ref{psiimage2})
and (\ref{psiimage}) will give the state of the propagated
ququarts

\begin{equation}        \label{Imqudits}
\ket{\Psi}_{Im} = \frac{1}{2} \sum_{l=-\frac{3}{2}}^{\frac{3}{2}}
          e^{ik\frac{d^{2}l^{2}}{2z_{A}}} \;
          \ket{-l}\lsub{1} \otimes \ket{l}\lsub{2},
\end{equation} which has the same form of the two-photon state at the plane of
the four-slits, Eq.~(\ref{qudits}), showing that this state can be
propagated at the free-space. A more important result can be
obtained when one uses Eq.~(\ref{psiimage}) and
Eq.~(\ref{psiimage2}) to show that the general state for the
twin-photons after being transmitted through generic apertures
(See Eq.~(\ref{psislits})), will always be reconstructed at the
planes of their images. It is not worthless to mention that the
theory doesn't require the use of identical lenses what means that
the receivers of the entangled qudits can be at different
distances from the apertures (source).


\section{Experimental setup and results}

\begin{figure}[tbh]
\begin{center}
\includegraphics[width=0.22\textwidth]{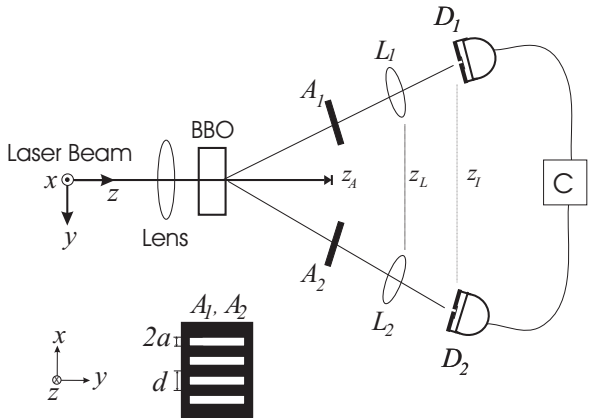}
\end{center}
\caption{\label{fig:setup}  Outline of the experimental setup.
$A_{1}$ and $A_{2}$ are four-slit apertures, $L_{j}$ lens, $D_{j}$
a detector and C is a coincidence counter.}
\end{figure}

The setup represented in Fig.~\ref{fig:setup} was carried out to
experimentally demonstrate the free-space propagation of the
ququarts entangled state described by Eq.~(\ref{qudits}). A
5-mm-long $\beta$-barium borate crystal is used to generate
type-II parametric down-conversion luminescence when pumped by a
100~mW pulsed laser beam. Down-converted photons with the same
wavelength ($\lambda = 826$~nm) are selected using interference
filters. Two identical four-slits ($A_{1}$ and $A_{2}$) are placed
at their exit path at the same distance $z_{A}= 200$~mm from the
crystal ($z=0$). The slit width is $2a\approx0.09$~mm and the
distance between two consecutive slits is $d\approx0.17$~mm. To
guarantee that the function $W(\xi ;z_{A})$ is non-vanishing only
at a small region of the space the pump beam transverse profile
was focused at the plane of these apertures. After being
transmitted by the four-slits the biphotons are propagated at the
free-space and through two identical lenses ($L_{1}$ and $L_{2}$),
with focal length of $f = 150$~mm and placed at a distance of
$z_{L}= 500$~mm from the crystal. At the plane of the image
formation of the apertures ($z_{I}= 800$~mm), avalanche photodiode
detectors ($D_{1}$ and $D_{2}$) are placed. Single and coincidence
counts are measured by the detectors and in front of each detector
there is a single slit of width 0.1~mm (parallel to the
four-slits).

Coincidence selective measurements onto the basis
$\{\ket{l}\lsub{1}\ket{l'}\lsub{2}\}$ are performed to determine
the two-photon image state. Detector $D_{1}$ is fixed at a region
in space where the image of slit $l$ of the four-slit in channel 1
is formed while detector $D_{2}$ is scanning, in the $x$
direction, the entire region where the image of the other
four-slit is formed. Four measurements of this kind, with detector
$D_{1}$ going from the image of the slit for which
$l=\frac{-3}{2}$ to the image of the slit with $l=\frac{3}{2}$,
will determine the probability amplitudes in the sixteen basis
states $\{\ket{l}\lsub{1}\ket{l'}\lsub{2}\}$.

If the theoretical result of Eq.~(\ref{Imqudits}) for the state of
the twin photons at the plane of image formation is correct,
coincidences peaks will occur only when detector $D_{2}$ passes by
the image of slit for which $l'=-l$. However, the classical
correlated state given by

\begin{equation}        \label{classico}
\rho_{\text{cc}} = \frac{1}{2} \sum_{l=-\frac{3}{2}}^{\frac{3}{2}}
       \ket{l}\lsub{1\,1\!}\!\bra{l} \otimes
       \ket{-l}\lsub{2\,2\!\!}\bra{-l},
\end{equation} predicts the same experimental result. And then to guarantee
that the image state is indeed given by a coherent superposition
(Eq.~(\ref{Imqudits})), the detectors are moved to a distance of
$200$~mm from the image formation plane and conditional
fourth-order interference patterns \cite{Greenberger,Fonseca} are
measured. As it was demonstrated in Ref. \cite{Glima}, when we
treat the spatial correlations of two photons, the observation of
a fourth-order interference pattern with conditional fringes is a
sufficient signature for entanglement. If the correlations between
the propagated ququarts were classical, the coincidence count
rate, at this new configuration of the setup, would exhibit only a
single slit diffraction pattern.

\begin{figure}[t]
\begin{center}
\includegraphics[width=0.40\textwidth]{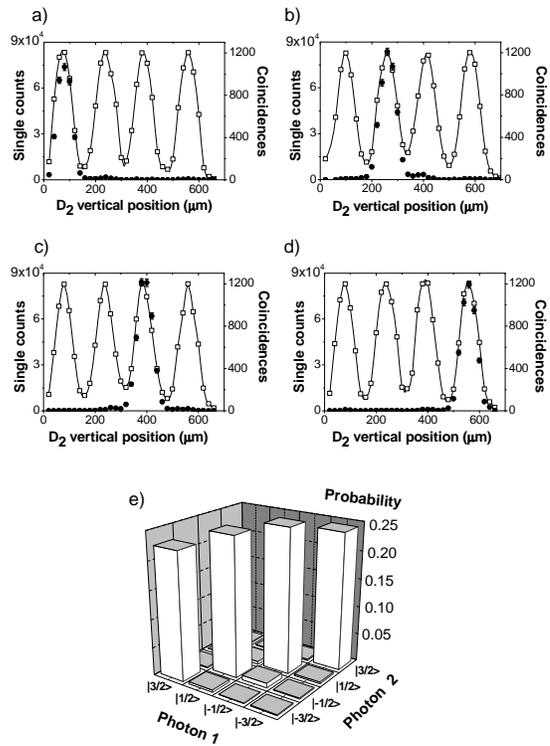}
\end{center}
\caption{\label{fig:ququart} $D_{2}$ single counts ($\Box$) and
coincidence counts ($\bullet$) measured in 20~s, simultaneously,
with $D_{1}$ fixed behind the image of the slit $l$ in channel 1
and $D_{2}$ scanning in $x$ direction the image of the four-slit
in channel 2. From left to right the single count peaks are the
slits $l'=-\frac{3}{2},\ldots,+\frac{3}{2}$.
 $D_{1}$ is fixed behind the slit $l$ (a) $\st +\frac{3}{2}$,
(b) $\st +\frac{1}{2}$, (c) $\st -\frac{1}{2}$ and (d) $\st
-\frac{3}{2}$. (e) 3D-Histogram of the probabilities measured for
all basis states $\{\ket{l}\lsub{1}\ket{l'}\lsub{2}\}$.}
\end{figure}

The experimental data recorded at the plane of image formation of
the four-slits is showed in Fig.~\ref{fig:ququart}. We can see
that the results are in agrement with Eq.~(\ref{Imqudits}) because
coincidences peaks were only observed when $D_{2}$ was scanning
the image of the slit symmetrically opposite to that which
detector $D_{1}$ was fixed. Figure~\ref{fig:ququart}(e) is a
histogram constructed using all the coincidences recorded in the
four measurements performed. The probability related is the chance
for the propagated ququarts state, selected in coincidence, be in
the form of one of the basis sates. The fact that the
probabilities for the states $\ket{l}\lsub{1}\ket{-l}\lsub{2}$ are
almost equally weighted and all the others probabilities null is a
strong evidence that the image state is a \emph{maximally}
entangled state of ququarts. This means that the states
$\ket{l}\lsub{1}\ket{-l}\lsub{2}$ will have almost the same
amplitudes at the coherent superposition of the obtained image
state (See Eq.~(\ref{expquarts})).

\begin{figure}[tbh]
\begin{center}
\includegraphics[width=0.30\textwidth]{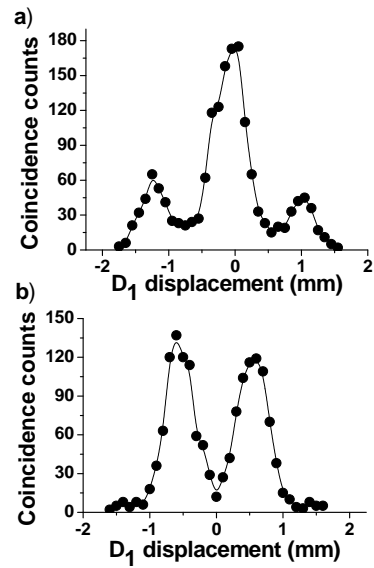}  
\end{center}
\caption{\label{interf}  Fourth-order interference patterns  as a
function of $D_{1}$ position. They were recorded when the
propagated photons were detected in coincidence at a distance of
$200$~mm of the image formation plane. (a) $D_{2}$ is kept fixed
at $x_{2}=0$~mm. (b) $D_{2}$ is kept fixed at $x_{2}=0.6$~mm. The
solid curve is a guide to the reader's eyes.}
\end{figure}

The fourth-order interference patterns measured when the detectors
were moved to a distance of $200$~mm from the image formation
plane and the propagated ququarts detected in coincidence are
showed in Fig.~\ref{interf}. Coincidence measurements were made as
a function of the $x$ position of the detector $D_{1}$ while
$D_{2}$ was kept fixed. The results are shown in
Fig.~\ref{interf}: (a) $D_{2}$ fixed at $x_{2}=0$~mm; (b) $D_{2}$
fixed at $x_{2}=0.6$~mm. The visibilities of the interference
patterns are $v_{a} = 0.86 \pm 0.05$ and $v_{b} = 0.83 \pm 0.04$,
respectively. One can easily observe the conditionality of the
fringes of these patterns. As mentioned before, the presence of
conditional interference patterns, at this new configuration of
the setup, demonstrates that the image state is pure and
entangled.

The propagated ququarts state obtained from the measurements can
then be written as:

\begin{eqnarray}
\ket{\Psi}_{Im} & = &   0.49\;\ket{\st -\frac{1}{2}, +\frac{1}{2}}
      \; + \;  0.50\;\ket{\st +\frac{1}{2}, -\frac{1}{2}} \nonumber \\
  &  & \!\!\!\!\!\!\!\! \text{} +
     e^{i\frac{kd^{2}}{z_{A}}} (0.47\;\ket{\st -\frac{3}{2}, +\frac{3}{2}}
      \; \!\!\! + \!\! \;  0.49\;\ket{\st +\frac{3}{2}, -\frac{3}{2}}),
 \label{expquarts}
\end{eqnarray} which has a fidelity of $F=0.98\pm 0.06$ to the original state of the
ququarts given by Eq.~(\ref{qudits}). This prove that we were able
to propagate the entangled ququarts state keeping a high-fidelity
to its original form. The phase in Eq.~(\ref{expquarts}) was not
measured because it can be cancelled out by choosing right values
for $d$ and $z_{A}$ or by adding an appropriate external phase to
a given slit.


\section{Discussion and conclusion}

We believe that the process of entangled qudits propagation,
described above, can be implemented for larger distances. As it
was demonstrated, at the plane of image formation of the
four-slits, the ququarts entangled state (Eq.~(\ref{qudits})) is
reconstructed with high-fidelity. Besides this, it is well know
that different configurations of lenses can be used after objects
to make their image appear at long distances. So, one can see that
the use of such configuration after the apertures would allow the
transmission of the entangled photons through more significant
distances. In Ref. \cite{Zeilinger1}, two telescopes were used to
propagate entangled qubits over more than $500$~m.

In conclusion, we have presented a principle to propagate
entangled states of qudits, generated using the transverse
correlation of the twin photons produced by SPDC, at the
free-space. Up to our knowledge this is the first report of a
propagation of entangled states of high-dimensional quantum
systems. The experimental test performed obtained a propagated
state with a high-fidelity to its original form. The benefits of a
free-space distribution of quantum entanglement were already
discussed at Ref. \cite{Zeilinger1}. The advantages of using
entangled states of high-dimensions quantum systems to transmit
information come both from the increase at the quantity of
information that can be encoded at the entangled quantum systems
and for the possibility of performing more safers quantum
cryptography protocols. For these reasons we believe that the work
presented in this paper is an important step that can be
considered in the effort of building quantum communication
networks.

\section{Acknowledgments}

This work was supported by the Brazilian agencies CAPES, CNPq,
Fapemig and Mil\^enio-Informa\c{c}\~ao Qu\^antica. C. Saavedra was
supported by Grants Nos. FONDECYT 1040591 and Milenio ICM P02-49F.

\end{document}